\documentclass[aps,prl,twocolumn,groupedaddress]{revtex4}
\usepackage{graphicx} 
\usepackage{dcolumn}
\usepackage{bm}
\begin{document}

\title{The Curie temperature and exchange energy between two
sublattices in half-metallic
greigite Fe$_{3}$S$_{4}$} 
\author{Jun Wang$^{1,a}$, Shi-He Cao$^{1}$, Wei Wu $^{1}$, and Guo-meng Zhao$^{1,2,b}$} 
\affiliation{
$^{1}$Department of Physics, Faculty of Science, Ningbo
University, Ningbo, P. R. China~\\
$^{2}$Department of Physics and Astronomy, 
California State University, Los Angeles, CA 90032, USA}

\begin{abstract}

High-temperature magnetic measurements have been carried out in hydrothermally 
synthesized greigite (Fe$_{3}$S$_{4}$). We show that the Curie
temperature of greigite is 
significantly lower than that for its iron oxide counterpart Fe$_{3}$O$_{4}$. 
The lower $T_{C}$ value (about 677 K) of greigite is in quantitative agreement with that
calculated using the exchange energy (3.25 meV) and the spin values of the two
sublattices, which are inferred from the neutron and magnetization data of
high-quality pure greigite samples. We further show that, with an effective on-site Hubbard energy $U_{eff}$ =
1.16 eV, the lattice constant and two sublattice spins predicted from {\em ab initio} 
density-function
theory are in nearly perfect agreement with the measured values.  
The parameter $U_{eff}$ = 1.16 eV ensures Fe$_{3}$S$_{4}$ to be
an excellent half-metallic material for spintronic applications. 
 
\end{abstract}
\maketitle 

Greigite (Fe$_{3}$S$_{4}$) that was first discovered 
in lake sediments from California \cite{Skinner} is an iron thiospinel. Greigite 
was found to have the
inverse spinel structure \cite{Skinner} like its iron oxide counterpart,
magnetite (Fe$_{3}$O$_{4}$).  Greigite has been used as a recorder of
the ancient geomagnetic field and environmental processes and is important for paleomagnetic and environmental
magnetic studies \cite{Snowball,Jiang,Bab}. It is also widespread 
in magnetotactic
bacteria that produce greigite magnetosomes \cite{Baz,Po,Kas}. Therefore, greigite is of 
general interest in geophysics and biology.   

On the other hand,  greigite has been less known to 
physicists and material scientists because pure greigite samples are 
difficult to synthesize and some of its fundamental
magnetic properties are still unknown. The precise value of the Curie
temperature $T_{C}$ remains unknown 
despite variable estimations since 1974. The difficulty in
determination of the Curie temperature is due to the fact that greigite is chemically 
unstable at high 
temperatures even in argon environment \cite{Dek}.
High-temperature magnetic measurements often
revealed chemical decomposition that precluded from determination of the Curie temperature
\cite{Dek,Robert}. Spender {\em et al.} \cite{Spender}
estimated $T_{C}$ to be 606 K by extrapolating
thermomagnetic curves to high temperatures. Vandenberghe {\em et al.} \cite{Van} 
made M\"ossbauer spectroscopic
measurements up to 480 K and extrapolated the effective field of iron in tetrahedral 
sites to obtain a 
$T_{C}$ value of at least 800 K. More recent magnetic measurements on 
high-quality pure greigite samples \cite{Chang08} suggested a
Curie temperature of higher than 630 K. Therefore, the reported Curie 
temperature of greigite ranges from 600 K to 800 K.

Another important magnetic property is the saturation magnetization
which is associated with the spin values of
(Fe$^{3+}$)$_{A}$ and (Fe$^{3+}$Fe$^{2+}$)$_{B}$ on the respective tetrahedral 
(A) and octahedral (B) sublattices.
 Coey {\em et al.} \cite{Coey} suggested that greigite should have a net magnetic moment 
 $m$ $>$ 4 $\mu_{B}$  per formula unit considering the spin-only values 
 for the ionic moments. They could not account for the measured magnetic moment
of 2.2 $\mu_{B}$.  Recent neutron diffraction experiments on  
high-quality pure greigite samples \cite{Chang09} indicate a much larger
net magnetic moment of 3.4~$\mu_{B}$. Devey {\em et
al.} \cite{Devey} studied the electronic and magnetic behaviors of this material using
{\em ab initio}
density-functional theory in the generalized gradient approximation
(GGA) with the on-site Hubbard energy $U_{eff}$
parameter (GGA + U).  They calculated the lattice constant $a$ and two sublattice
moments ($m_{A}$ and $m_{B}$) as a
function of $U_{eff}$. Comparing the measured lattice constant with their calculated 
results \cite{Devey}, 
they found that $U_{eff}$ $\simeq$ 1 eV. This $U_{eff}$ value leads to
$m_{A} \simeq m_{B} \simeq$~3$\mu_{B}$, in quantitative agreement the
measured values from neutron diffraction \cite{Chang09}.

Here we report high-temperature magnetic measurements on hydrothermally 
synthesized greigite. We show that the Curie
temperature of greigite is 
significantly lower than that for its iron oxide counterpart Fe$_{3}$O$_{4}$. 
The lower $T_{C}$ value (about 677 K) of greigite is in quantitative agreement with that
calculated using the exchange energy (3.25 meV) and the spin values of the two
sublattices, which are inferred from the neutron \cite{Chang09} and magnetization
\cite{Chang08} data of
high-quality pure greigite samples. We further show that, with an effective on-site Hubbard energy $U_{eff}$ =
1.16 eV, the lattice constant and two sublattice spins predicted from {\em ab initio} 
density-function
theory \cite{Devey} are in quantitative agreement with the measured values.  
The parameter $U_{eff}$ = 1.16 eV ensures Fe$_{3}$S$_{4}$ to be
an excellent half-metallic material for spintronic applications.

\begin{figure}[htb]
     \vspace{-0.2cm}
    \includegraphics[height=6cm]{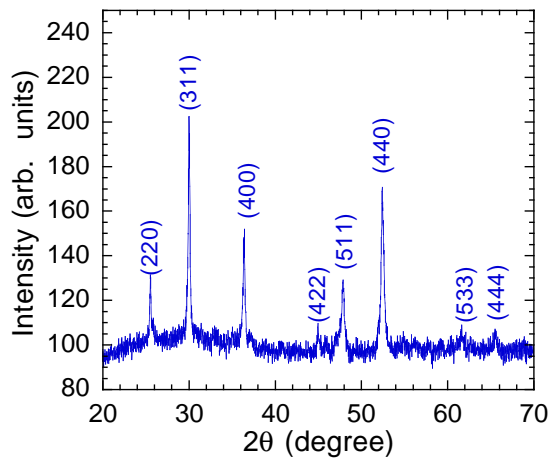}
     \vspace{-0.3cm}
 \caption[~]{X-ray diffraction (XRD) spectrum for a greigite sample synthesized under a magnetic 
field of 2.5 kOe. All the peaks can be indexed by the 
 spinel Fe$_{3}$S$_{4}$ phase with a lattce constant $a$ = 9.877~\AA. The mean diameter of the crystallites (subparticles) is determined to be
about 36 nm from the width of the (311) peak.} 
\end{figure}

Greigite samples were
synthesized under a magnetic field with a hydrothermal route \cite{He}.  
Briefly, ammonium iron sulfate hexahydrate (0.2 mmol) and 
cysteine (1 mmol) were dissolved in distilled water (25 mL) and stirred 
vigorously for 50 min. The solution was then put into a 30 mL Teflon vessel with 
two circular magnets (2.5 kOe) at the top and bottom in a stainless steel autoclave. After 12 h 
reaction at 180 $^{\circ}$C, black products were obtained. These products were washed 
several times in distilled water and pure ethanol to remove impurities. These samples 
were dried in vacuum at 60 $^{\circ}$C for at least 6 h.

Samples were checked by x-ray diffraction (XRD)
right after synthesis. 
Fig.~1a shows XRD spectrum for a greigite sample synthesized under a magnetic 
field of 2.5 kOe. All the 
peaks can be indexed by the spinel Fe$_{3}$S$_{4}$ phase. The lattce constant $a$ 
is evaluated to be 9.877~\AA, in excellent
 agreement with the reported value (9.876~\AA) \cite{Skinner,Chang09}. The mean 
 diameter of the crystallites (subparticles) is determined to be
about 36 nm from the width of the (311) peak. 

Magnetization was measured using a Quantum Design vibrating sample magnetometer (VSM).
The moment measurement was carried out after the sample chamber reached a high vacuum of 
better than 9$\times$10$^{-6}$ torr. The absolute measurement uncertainties 
in temperature and moment are less than 10 K and 1$\times$10$^{-6}$
emu, respectively. We used the same heating and cooling rate of 30
K/min for thermomagnetic measurements, which can ensure a thermal lag of 
less than 10 K.

In Fig.~2, we plot magnetic hysteresis loop at
300 K for the greigite sample. Since the magnetization is almost saturated 
at a magnetic field of
20 kOe, we can take the magnetization at 20 kOe to be the saturation 
magnetization $M_{s}$. The 
$M_{s}$ value is found to be about 40 emu/g, which is close to that of the similarly prepared
sample \cite{He} but significantly below the value (59 emu/g) reported for 
high-quality greigite samples with $d$ = 14 $\mu$m (Ref.~\cite{Chang08}). The reduction 
of the $M_{s}$
value in our sample may be due to both finite-size effects \cite{Res} and
possible presence
of nonmagnetic amorphous phases.

\begin{figure}[htb]
     \vspace{-0.2cm}
    \includegraphics[height=6cm]{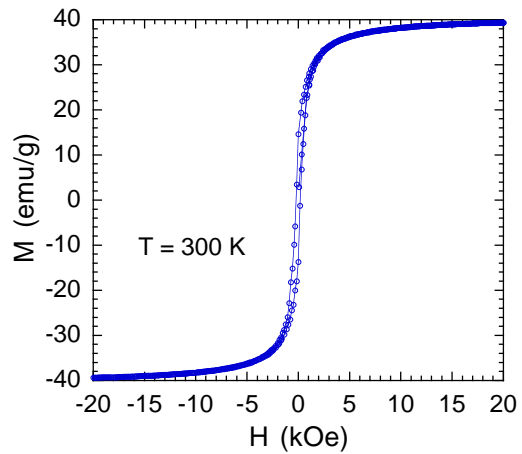}
     \vspace{-0.3cm}
 \caption[~]{Magnetization versus magnetic field 
at 300 K. The saturation magnetization $M_{s}$ is found
to be about 
40 emu/g.
} 
\end{figure}

Figure 3a shows the temperature dependence of the magnetization for
the greigite sample, which was measured in a magnetic field of 10 kOe. The warm-up data up to 
780 K indicate a magnetic transition around 700 K. A substantial magnetization 
between 720 and 780 K implies that a small fraction of the Fe$_{3}$S$_{4}$
phase was converted to the magnetic Fe$_{3}$O$_{4}$ phase in this
temperature regime. Since this sample was measured after the VSM high-vacuum cryopump 
was exposed to air during a service and 
regenerated through evacuation by a mechanic pump, it is likely that
there is a minor oxygen contamination that may promote a chemical alteration
to the Fe$_{3}$O$_{4}$ phase below 700 K.

\begin{figure}[htb]
     \vspace{-0.2cm}
    \includegraphics[height=12cm]{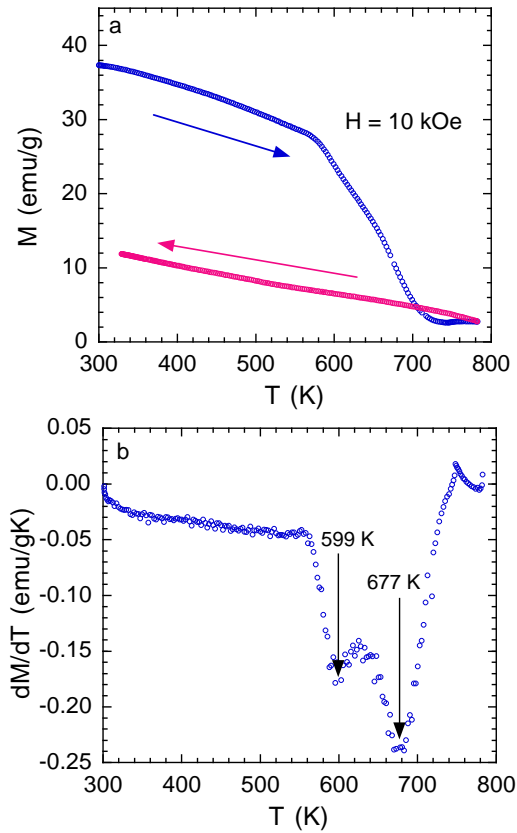}
     \vspace{-0.5cm}
 \caption[~]{a) Temperature dependence of the magnetization for sample
 A, which was measured in a field of 10 kOe. b) Temperature dependence 
 of the derivative of the warm-up magnetization.} 
\end{figure}

After the sample was cooled down from 780 K with a rate of 30 K/min, the magnetic
transition around 700 K disappeared and the magnetization at room
temperature was reduced to about 30$\%$ of the initial value,
indicating that the Fe$_{3}$S$_{4}$ phase was completely decomposed into other
phases (e.g., Fe$_{3}$O$_{4}$ and nonmagnetic
ferrosulfides such as FeS$_{2}$ \cite{Dek}) after the high-temperature magnetic measurement. 

In order to see the magnetic transition more clearly, we plot
the derivative ($dM/dT$) of the warm-up magnetization with respective to temperature. 
It is apparent that a minimum in $dM/dT$ occurs at a temperature of
about 677 K. The temperature corresponding to the minimum in $dM/dT$ or the inflection point of the
$M (T)$ curve should be the Curie temperature of greigite, that is, $T_{C}$
$\simeq$ 677 K. There is also a second local minimum at about 599 K,
which happens to the same as the Curie temperature of Fe$_{7}$S$_{8}$
(Ref.~\cite{Dek89}). This suggests that a small fraction of the Fe$_{3}$S$_{4}$ phase
was converted to the Fe$_{7}$S$_{8}$ phase even at a temperature below
599 K. Since the oxygen contamination in our VSM system is so minor
and the warming rate is high (30 K/min), the remaining Fe$_{3}$S$_{4}$ phase is still 
substantial when the Curie temperature of the phase is reached. This allows us to 
determine the Curie temperature of Fe$_{3}$S$_{4}$ quite reliably.
The magnetic transition at 677 K should be associated
with the ferrimagnetic transition of greigite unless Fe$_{3}$S$_{4}$ would 
have been converted to an unknown magnetic iron sulfide phase.

\begin{figure}[htb]
     \vspace{-0.2cm}
    \includegraphics[height=13.0cm]{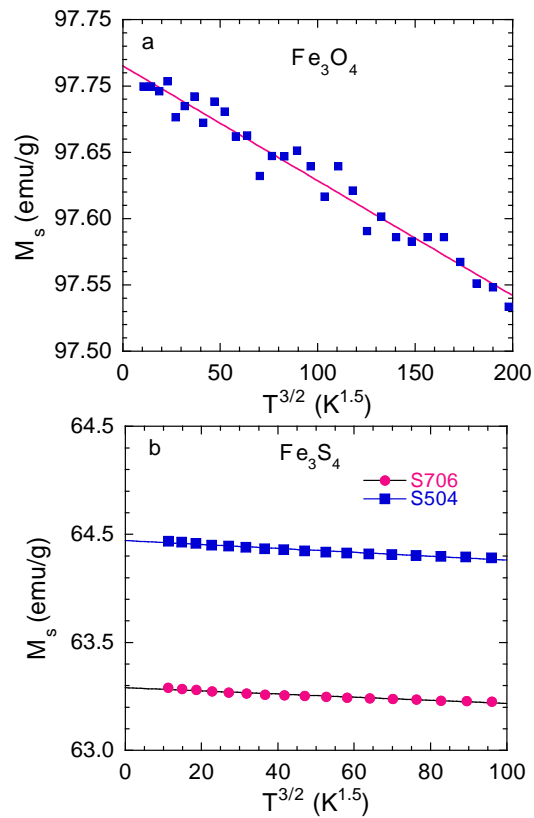}
     \vspace{-0.6cm}
 \caption[~]{a) Low-temperature saturation magnetization versus
 $T^{3/2}$ for a Fe$_{3}$O$_{4}$ crystal. The data are extracted from
 Ref.~\cite{Ara}. b) Low-temperature saturation magnetizations versus
 $T^{3/2}$ for two Fe$_{3}$S$_{4}$ samples (S706 and S504). The data
 are taken from Ref.~\cite{Chang08}.} 
\end{figure}

Our data suggest that the intrinsic Curie temperature 
of greigite is about 677 K. The intrinsic net moment of greighte  has 
been determined to be about 3.4 $\mu_{B}$ per 
formula unit from both magnetization \cite{Chang08} and neutron
\cite{Chang09} data. Both the Curie temperature and 
the magnetic moment of greigite are significantly lower than 
those of its counterpart Fe$_{3}$O$_{4}$. 

In order to quantitatively understand the
differences between the two spinels, we need to consistently extract the exchange energy $J_{AB}$ 
between the A and B sublattices for these two
compounds. One way to determine $J_{AB}$ is from the temperature dependence of the 
saturation (or high-field) magnetization at low temperatures.  Since exchange 
energies ($J_{AA}$ and $J_{BB}$) between
ions in the same sublattice of the inverse spinel structure are
negligibly small \cite{Glass}, we can neglect these exchange energies 
and only keep $J_{AB}$. Within this approximation, the
dispersion relation for the acoustic magnon to order $k^{2}$ is 
\cite{Glass}:
\begin{equation}
    \hbar\omega =
    \frac{22}{16}\frac{J_{AB}S_{A}S_{B}}{2S_{B}-S_{A}}k^{2}a^{2},
    \end{equation}
    where $\omega$ is the frequency and $k$ is the wave-number of the
    spin wave. From this dispersion, one can readily obtain a relation for the
    saturation magnetization at low temperatures: $M_{s}(T) = M_{s} (0)
    (1 - CT^{3/2})$ with \cite{Ara}
    \begin{equation}
    C =
    \frac{0.05864}{4(2S_{B}-S_{A})}[\frac{16(2S_{B}-S_{A})k_{B}}{22J_{AB}S_{A}S_{B}}]^{3/2},
    \end{equation}
    where $k_{B}$ is the Boltzman's constant.
 
In Figure 4, we plot low-temperature saturation magnetizations versus
 $T^{3/2}$ for a Fe$_{3}$O$_{4}$ crystal (Fig.~4a) and for two Fe$_{3}$S$_{4}$ samples
 (Fig.~4b). The data for the Fe$_{3}$O$_{4}$ crystal are extracted from
 Ref.~\cite{Ara} and the data for two Fe$_{3}$S$_{4}$ samples are from
Ref.~\cite{Chang08}. The best linear fits to the data yield $C$ =
(8.9$\pm$0.3)$\times$10$^{-6}$ K$^{-1.5}$ for Fe$_{3}$O$_{4}$; $C$ =
(1.15$\pm$0.06)$\times$10$^{-5}$ K$^{-1.5}$ for sample S706; $C$ =
(1.42$\pm$0.07)$\times$10$^{-5}$ K$^{-1.5}$ for sample S504. The
average $C$ value for Fe$_{3}$S$_{4}$ is calculated to be 1.29$\times$10$^{-5}$ K$^{-1.5}$.

For Fe$_{3}$O$_{4}$, $S_{A}$ = 2.5 and  $S_{B}$ = 2.25 (Ref.~\cite{Glass}). Substituting
$C$ = (8.9$\pm$0.3)$\times$10$^{-6}$ K$^{-1.5}$, $S_{A}$ = 2.5, and $S_{B}$ = 2.25
into Eq.~(2) yields $J_{AB}/k_{B}$ = 22.7$\pm$0.6 K, which is slightly
below the value (27.8 K) determined directly from neutron scattering \cite{Glass}. For
Fe$_{3}$S$_{4}$, the two sublattice moments have also been determined 
by neutron diffraction \cite{Chang09}. From the measured values
\cite{Chang09}: 
$m_{A}$ = 3.08 $\mu_{B}$ and $m_{B}$ = 3.25 $\mu_{B}$, we find $S_{A}$ =
1.54 and  $S_{B}$ = 1.63. Substituting
$S_{A}$ = 
1.54, $S_{B}$ = 1.63, and the average $C$ value of 1.29$\times$10$^{-5}$ K$^{-1.5}$ 
into Eq.~(2), we find $J_{AB}/k_{B}$ = 37.7 K. 

With neglect of exchange interactions between
ions in the same sublattice of the inverse spinel structure, the Curie temperature  
within a mean-field approximation is given by \cite{Kou}:

\begin{equation}
    T_{C}^{MF} =
    4\sqrt{2}(J_{AB}/k_{B})\sqrt{S_{A}S_{B}(S_{A}+1)(S_{B}+1)}
    \end{equation}
    
Since the mean-field theory tends to overestimate the Curie
temperature. Corrections to the mean-field theory were given
for different structures such as simple cubic (sc), body-centered
cubic (bcc), and face-centered cubic (fcc) \cite{Fisher}. Since the magnetic structure for
the inverse spinel is fcc \cite{Glass}, the theoretical Curie temperature $T_{C}$ 
after the correction to the mean-field theory is \cite{Fisher}:
\begin{equation}
    T_{C} = 0.789[1-0.13/\sqrt{S_{A}S_{B}(S_{A}+1)(S_{B}+1)}]T_{C}^{MF}.
    \end{equation}

For Fe$_{3}$O$_{4}$, $S_{A}$ = 2.5, $S_{B}$ = 2.25, and $J_{AB}/k_{B}$ = 22.7$\pm$0.6 K.
Substituting these numbers into Eqs.~(3) and (4), we find that $T_{C}$
= 808$\pm$22 K, which is slightly below the measured $T_{C}$ = 851 K
for Fe$_{3}$O$_{4}$ (Ref.~\cite{Ara}). For Fe$_{3}$S$_{4}$, $S_{A}$ = 
1.54, $S_{B}$ = 1.63, and $J_{AB}/k_{B}$ = 37.7 K. Pluging these 
values into Eqs.~(3) and (4) yields $T_{C}$
= 667 K, slightly below the measured value of 677 K. This quantitative 
agreement suggests that the sublattice moments obtained from the
neutron diffraction \cite{Chang09} and the Curie temperature measured in this work
represent the intrinsic magnetic properties of greigite.

Now we compare the first principle
calculations \cite{Devey} with the experimental results. Comparing the
measured lattice constant $a$ = 9.8538~\AA~at 10 K
(Ref.~\cite{Chang09})
with the calculated results \cite{Devey}
yields $U_{eff}$ = 1.16 eV, $m_{A}$ = 3.05~$\mu_{B}$, and  $m_{B}$ =
3.26~$\mu_{B}$. The predicted sublattice moments are in quantitative
agreement with the measured values \cite{Chang09}: $m_{A}$ = 3.08$\pm$0.08~$\mu_{B}$ and $m_{B}$ =
3.25$\pm$0.08~$\mu_{B}$. This quantitative agreement 
implies that the first principle calculations based on the GGA + U
model is very reliable. This also implies that the inferred parameter $U_{eff}$ =
1.16 eV is intrinsic to Fe$_{3}$S$_{4}$. This parameter  $U_{eff}$ =
1.16 eV is very important since it ensures greigite to be a half 
metal \cite{Devey}. Compared with other half metals such as CrO$_{2}$
(Ref.~\cite{Ji}) and doped
manganites \cite{Park,Wei,Zhao}, greigite should be 
better for
spintronic applications because it has a much higher Curie temperature than
those of CrO$_{2}$ and doped manganites. Compared with the
half-metallic magnetite (Ref.~\cite{Ver}),  Fe$_{3}$S$_{4}$
is also better for spintronic applications because it exhibits a highly conductive
metallic behavior \cite{Noz} in contrast to semiconductor-like electrical
transport in Fe$_{3}$O$_{4}$ \cite{Roz}. Therefore, greigite is an
excellent half-metallic material for spintronic applications.

In summary, we have determined the intrinsic Curie
temperature (677 K) and exchange energy (3.25 meV) between the two sublattices of greigite.  
The mesured Curie temperature is in quantitative agreement with that
calculated using the exchange energy and the spin values of the two
sublattices, which are independently determined from the neutron and magnetization data of
high-quality pure greigite samples \cite{Chang08,Chang09}. We further show
that, with an effective on-site Hubbard energy $U_{eff}$ =
1.16 eV, the lattice constant and two sublattice spins predicted from {\em ab initio} 
density-function
theory are in nearly perfect agreement with the measured values.  
The parameter $U_{eff}$ = 1.16 eV ensures Fe$_{3}$S$_{4}$ to be
an excellent half-metallic material for spintronic applications. The current work will
benefit to scientists working in 
multiple disciplines including physics, materials science
and technology, geophysics, geochemistry, and biology.

{\bf Acknowledgment:}
This work was supported by the National Natural Science Foundation of China (10874095), 
the Science Foundation of China, Zhejiang (Y407267, 2009C31149), the Natural Science Foundation of Ningbo 
(2008B10051, 2009B21003), K. C. Wong Magna Foundation, and Y. G. Bao's Foundation. 

~\\
~\\
$^{a}$ wangjun2@nbu.edu.cn~\\
$^{b}$ gzhao2@calstatela.edu

\bibliographystyle{prsty}

\end{document}